\pdfoutput=1
\documentclass[10pt,conference]{IEEEtran}

\usepackage[x11names, svgnames, rgb, usenames,dvipsnames]{xcolor}
\usepackage{amsmath,amssymb,amsfonts,amsthm, mathtools}
\usepackage[pdftex]{graphicx}

\usepackage[T1]{fontenc}
\usepackage[utf8]{inputenc}

\usepackage{csquotes}
\usepackage{nicefrac}
\usepackage[textsize=tiny]{todonotes}
\usepackage{booktabs}
\usepackage[hyphens,spaces,obeyspaces]{url}
\usepackage[bookmarks=false]{hyperref}
\usepackage{tikz}

\usetikzlibrary{shapes.geometric, positioning, arrows, trees, quotes}
\usepackage{yquant}
\usepackage{physics}
\usepackage{csvsimple-l3}

\makeatletter
\let\MYcaption\@makecaption
\makeatother
\usepackage[font=footnotesize]{subcaption}
\makeatletter
\let\@makecaption\MYcaption
\makeatother

\usepackage[style=ieee, maxnames=6,minnames=1,date=year,doi=false,isbn=false,backend=biber, sortcites]{biblatex}
\usepackage[binary-units=true, detect-all=true]{siunitx}

\usepackage{nicematrix}
\usepackage{flushend}

\newtheorem{example}{Example}

\definecolor{Blue01}{rgb}{0.26,0.39,0.85}
\definecolor{Yellow12}{rgb}{1.0,0.88,0.1}
\definecolor{Gray23}{rgb}{0.66,0.66,0.66}

\begin{document}

\title{\LARGE Limiting the Search Space in Optimal Quantum Circuit Mapping\vspace*{-0.0em}}

\author{
	\IEEEauthorblockN{Lukas Burgholzer\IEEEauthorrefmark{1}\hspace*{1.5cm}Sarah Schneider\IEEEauthorrefmark{1}\hspace*{1.5cm}Robert Wille\IEEEauthorrefmark{1}\IEEEauthorrefmark{2}}
	\IEEEauthorblockA{\IEEEauthorrefmark{1}Institute for Integrated Circuits, Johannes Kepler University Linz, Austria}
	\IEEEauthorblockA{\IEEEauthorrefmark{2}Software Competence Center Hagenberg GmbH (SCCH), Austria}
	\IEEEauthorblockA{\href{mailto:lukas.burgholzer@jku.at}{lukas.burgholzer@jku.at}\hspace{1.5cm}\href{mailto:sarah.schneider@jku.at}{sarah.schneider@jku.at} \hspace{1.5cm}\href{mailto:robert.wille@jku.at}{robert.wille@jku.at}\\
	\url{https://iic.jku.at/eda/research/quantum/}}
	\vspace*{-2em}
}

\maketitle

\begin{abstract}
	Executing quantum circuits on currently available quantum computers requires compiling  
	them to a representation that conforms to all restrictions imposed by the targeted architecture.
	Due to the limited connectivity of the devices' physical qubits, an important step in the compilation process is to \emph{map} the circuit in such a way that all its gates are executable on the hardware.
	Existing solutions delivering optimal solutions to this task are severely challenged by the exponential complexity of the problem. 
	In this paper, we show that the search space of the mapping problem can be limited drastically while still preserving optimality.
	The proposed strategies are generic, \mbox{architecture-independent}, and can be adapted to various mapping methodologies.
	The findings are backed by both, theoretical considerations and experimental evaluations. Results confirm that, by limiting the search space, optimal solutions can be determined for instances that timeouted before or speed-ups of up to three orders of magnitude can be achieved.  
\end{abstract}

\section{Introduction}\label{sec:introduction}

Capabilities of existing quantum computers are steadily growing, as, e.g., witnessed by IBM's hardware roadmap revealed at the end of 2020 that predicted a device with more than a thousand qubits for the year 2023~\cite{gambettaIBMRoadmapScaling2020}.
While these devices will most certainly not be able to run any useful instances of the famous Grover algorithm~\cite{groverFastQuantumMechanical1996} or Shor's algorithm~\cite{shorPolynomialtimeAlgorithmsPrime1997}, Variational Quantum Algorithms~\cite{cerezoVariationalQuantumAlgorithms2020} (besides many others) have been proposed as a promising way for actually making use of near-term quantum computers, e.g., by using the VQE algorithm to determine the ground state energy of a molecule~\cite{peruzzoVariationalEigenvalueSolver2014,kandalaHardwareefficientVariationalQuantum2017}.

In order to execute a conceptual quantum algorithm on an actual device, the algorithm or circuit has to be \emph{compiled} to a representation that adheres to all constraints imposed by the targeted device.
Since quantum computers typically only support a limited set of elementary operations, the algorithm's description first has to be \emph{decomposed} to the corresponding gate set~\cite{barencoElementaryGatesQuantum1995, maslovAdvantagesUsingRelative2016,willeImprovingMappingReversible2013,degriendArchitectureawareSynthesisPhase2020}.
In addition, not all physical qubits on the device might directly interact with each other.
As a consequence, the resulting circuit additionally has to be \emph{mapped} to the architecture of the device, so that any two-qubit operation is applied to physical qubits connected on the device.
Typically, this is accomplished by inserting \textit{SWAP} operations that allow to \emph{swap} the circuit's 
logical qubits from one position to a connected one. Chaining those operations allows for arbitrary changes in position. 
However, since each added operation decreases the fidelity of the result when executed on the quantum computer, it is vital to keep the overhead of the resulting (mapped) circuit as low as possible.

Unfortunately, the search space that needs to be considered when determining the best possible mappings increases exponentially with the number of involved qubits.
Thus, many existing solutions trade off accuracy/minimality for speed~\mbox{\cite{zulehnerEfficientMethodologyMapping2019, smithQuantumComputationalCompiler2019, liTacklingQubitMapping2019, matsuoReducingOverheadMapping2019, muraliNoiseadaptiveCompilerMappings2019,amyStaqFullstackQuantum2019,sivarajahKetRetargetableCompiler2020, zulehnerCompilingSUQuantum2019, hirataEfficientConversionQuantum2011}}.
While these methods are constantly improving, it has been shown that there is lots of room for improvement as they frequently stray far from the optimum~\cite{willeMappingQuantumCircuits2019}.
Accordingly, several approaches for generating \emph{optimal} circuit mappings have been proposed 
in the recent past~\cite{siraichiQubitAllocation2018,zhangTimeoptimalQubitMapping2021, tanOptimalLayoutSynthesis2020,dealmeidaFindingOptimalQubit2019,zhuExactQubitAllocation2020, willeMappingQuantumCircuits2019}. 
However, these methods extensively explore the immense search space of the mapping problem---significantly limiting their efficiency and applicability.

In this work, we show that this search space can be drastically limited, \emph{while still guaranteeing optimal results}. 
More precisely,
we present generic, architecture-independent observations that allow to substantially reduce the number of permutations to be considered in front of each gate---the origin of the huge search space and complexity.
The key idea is that it suffices to permute just enough that any two qubits of the architecture may interact with each other.
Those observations are additionally backed by theoretical considerations (based on group theory) showing that corresponding limitations of the search space  are indeed guaranteed to preserve optimality. Based on that, strategies are proposed how these findings can be utilized in existing approaches for optimal quantum circuit mapping. 

Experimental evaluations 
confirm the resulting benefits. 
By limiting the search space using the strategies  
proposed in this work, instances that previously suffered from timeouts  
can now be mapped within minutes or speed-ups of up to three orders of magnitude can be achieved---\emph{all while preserving optimality}.
The proposed strategies have been integrated on top of the quantum circuit mapping tool QMAP, which is publicly available at~\url{https://github.com/iic-jku/qmap} as part of the open-source JKQ toolkit for quantum computing~\cite{willeJKQJKUTools2020}.

The remainder of this work is structured as follows: 
In \autoref{sec:background}, we provide a review of the mapping problem and what constitutes an optimal solution to this problem.
\autoref{sec:general} describes our observations that allow to reduce the number of permutations to be considered in front of every gate. 
Based on that, we afterwards back those observations with a theoretical consideration in Section~\ref{sec:formal}.
In Section~\ref{sec:heuristics}, we then propose strategies how these findings can be utilized in existing methods.
Experimental evaluations confirming the resulting benefits are summarized in \autoref{sec:results}, before \autoref{sec:conclusions} concludes the paper.

\section{Background}\label{sec:background}
In order to keep this work self-contained, this section establishes the necessary background on the quantum circuit mapping problem and, afterwards, reviews the main idea how existing optimal solutions address this problem. We refer the interested reader to the provided references for further details.

\subsection{Quantum Circuit Compilation and the Mapping Problem}\label{sec:mapping}

In order to execute a quantum circuit on an actual quantum computer, it needs to be compiled to a representation that conforms to all the constraints imposed by the architecture of the device. 
First, the quantum circuit must be expressed using elementary operations supported by the device---a step often referred to as \emph{decomposition} or \emph{synthesis}~\cite{barencoElementaryGatesQuantum1995, maslovAdvantagesUsingRelative2016,willeImprovingMappingReversible2013,degriendArchitectureawareSynthesisPhase2020}.
In the following, we assume the elementary gate set to consist of arbitrary single-qubit gates and the controlled-NOT operation (as, e.g., provided by IBM's quantum computers), since this constitutes the de-facto standard to date.
However, the findings in this work can readily be extended to alternative (future) gate-sets, e.g., including gates acting on more than two qubits.
 
In addition, most existing quantum computers (those based on superconducting qubits) have a rather limited connectivity between their qubits---typically described by a \emph{coupling graph}.
As a consequence, it is necessary to map a circuit's logical qubits (denoted $q_0,\dots,q_{n-1}$ in the following) to the device's physical qubits (denoted $p_0,\dots,p_{m-1}$ in the following) so that any elementary operation is applied to qubits connected on the device.
Only the most trivial quantum circuits can be directly mapped to the physical architecture.
In most cases, the mapping has to change dynamically throughout the circuit in order to conform to all the constraints.
This can be accomplished by using \textit{SWAP} gates that allow to interchange the position of two logical qubits on the architecture.

The \emph{mapping problem} (sometimes also synonymously referred to as \emph{qubit routing}, \emph{qubit placement}, or \emph{qubit allocation}), as it is considered in this work, describes the task of determining a representation of an \mbox{$n$-qubit} quantum circuit $G=g_0,\dots,g_{|G|-1}$ that conforms to all constraints imposed by an $m$-qubit architecture (described by a coupling graph $(V, E)$)---all while keeping the overhead of mapping the circuit, i.e., the number of added gates, as small as possible\footnote{There exist other objectives besides minimizing the number of added gates, e.g., minimizing execution \emph{time/depth} or maximizing execution \emph{fidelity}. As will be evident later on, the findings in this work can easily be extrapolated to these objectives by adequately adjusting the considered objective function.}.
When minimizing the number of added gates, we can restrict $G$ to only consists of two-qubit \textit{CNOT} gates since single-qubit gates are not affected by the limited connectivity and, thus, do not require mapping.

\begin{figure}[t]
    \centering
    \begin{subfigure}[t]{0.29\linewidth}
 \centering
 \resizebox{0.9\linewidth}{!}{
 \begin{tikzpicture}
 	\begin{yquant}
		qubit {$\reg_{\idx}$} q[4];
		cnot q[3] | q[0];
		cnot q[3] | q[1];	
		cnot q[2] | q[0];	
		cnot q[1] | q[0];	
 	\end{yquant}
 \end{tikzpicture}}
 \caption{Quantum circuit $G$}\label{fig:sample}
    \end{subfigure}
    \hfill    
    \begin{subfigure}[t]{0.30\linewidth}
 \centering
 \resizebox{0.5\linewidth}{!}{
 \begin{tikzpicture}[
			roundnode/.style={circle, draw=green!60, fill=green!5, very thick, minimum size=7mm},
			swapnode/.style={circle, draw=red!60, fill=red!5, very thick, minimum size=5mm},
			]
			\node[roundnode] (p0)  {$p_0$};
			\node[roundnode] (p1)  [above=of p0] {$p_1$};
			\node[roundnode] (p2)  [right=of p1] {$p_2$};
			\node[roundnode] (p3)  [below=of p2] {$p_3$};
			
			\draw[-] (p0) -- (p1);
			\draw[-] (p1) -- (p2);
			\draw[-] (p2) -- (p3);
		\end{tikzpicture}}
 \caption{$4$-qubit architecture}\label{fig:linear_architecture}
    \end{subfigure}
        \hfill    
    \begin{subfigure}[t]{0.38\linewidth}
 \centering
 \resizebox{0.99\linewidth}{!}{
 \begin{tikzpicture}
 	\begin{yquant}
 	qubit {$q_1 \mapsto \reg_{0}$} p[1];
 	qubit {$q_3 \mapsto \reg_{1}$} p[+1];
 	qubit {$q_0 \mapsto \reg_{2}$} p[+1];
 	qubit {$q_2 \mapsto \reg_{3}$} p[+1];

 	cnot p[1] | p[2];
 	cnot p[1] | p[0];
	 swap (p[0],p[1]);
	 align p[1], p[2];
 	cnot p[3] | p[2];	
 	cnot p[1] | p[2];	
 	\end{yquant}
 \end{tikzpicture}}
 \caption{Mapped circuit $G'$}\label{fig:mapped_circuit}
    \end{subfigure}
    \caption{Quantum circuit, architecture, and potential mapping}\label{fig:circ_arch}
    \vspace*{-1.5em}
\end{figure}
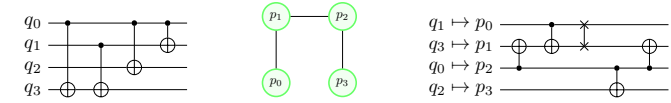

\begin{example}
	Consider the four-qubit quantum circuit~$G$ composed of four \textit{CNOT} gates as shown in \autoref{fig:sample} and assume it shall be mapped to a four-qubit (linear) architecture described by the coupling graph
	\mbox{$
	(V, E) = (\{p_0, p_1, p_2, p_3\}, \{e_{01}, e_{12}, e_{23}\})
	$}, which is shown in~\autoref{fig:linear_architecture}.
	Then, \autoref{fig:mapped_circuit} shows one possible mapping of $G$ to this architecture.
	By assigning $q_0\mapsto p_2$, $q_1 \mapsto p_0$, $q_2\mapsto p_3$, and $q_3 \mapsto p_1$, only a single \textit{SWAP} operation applied to $p_0$ and $p_1$ is needed in order for all gates to be executable.
\end{example}

While many (heuristic) techniques have been proposed in the past that allow to determine suitable mappings, e.g., ~\mbox{\cite{zulehnerEfficientMethodologyMapping2019, smithQuantumComputationalCompiler2019, liTacklingQubitMapping2019, matsuoReducingOverheadMapping2019, muraliNoiseadaptiveCompilerMappings2019,amyStaqFullstackQuantum2019,sivarajahKetRetargetableCompiler2020, hirataEfficientConversionQuantum2011, zulehnerCompilingSUQuantum2019}}, determining truly optimal solutions (with as little overhead as possible) revealed to be a challenging problem.
In fact, the mapping problem has been shown to be NP-complete~\cite{boteaComplexityQuantumCircuit2018, siraichiQubitAllocation2018}.

\subsection{Optimal Solutions for the Mapping Problem}\label{sec:sat}

The complexity of the mapping task mainly comes from the fact that, in principle, any possible permutation of the logical qubits (eventually realized as a series of \mbox{architecture-conforming} \textit{SWAP} operations) might be applied in front each gate of the circuit in order to realize a conforming mapping.
An optimal solution to the mapping problem can be determined by finding the right permutations to apply in front of every gate so that the overall resulting number of \textit{SWAPs} is minimal.
As a result, for a circuit $G$ with $|G|$ gates to be mapped to an \mbox{$m$-qubit} architecture the search space comprises a total of $|G|*m!$ permutations\footnote{
It has been shown in~\cite{willeMappingQuantumCircuits2019} that grouping of gates, e.g., to capture parallel execution of gates, is not guaranteed to produce gate-optimal results. Hence, it is indeed necessary to consider an arbitrary permutation in front of every single gate in order to achieve gate-optimal results.}.

\begin{figure}[t]
	
	\resizebox{\linewidth}{!}{
	\begin{tikzpicture}
	\begin{yquant}
		qubit {$q_\idx \mapsto \reg_{\idx}$} p[4];
		box {$\pi_0\in\Pi$} (p);
		[draw=none]
      	box {$p_{\pi_0(0)}$} p[0];
      	[draw=none]
      	box {$p_{\pi_0(1)}$} p[1];
      	[draw=none]
      	box {$p_{\pi_0(2)}$} p[2];
      	[draw=none]
      	box {$p_{\pi_0(3)}$} p[3];
		[name=g0]
		cnot p[3] | p[0];
		box {$\pi_1\in\Pi$} (p);
		[draw=none]
      	box {$p_{\pi_1(\pi_0(0))}$} p[0];
      	[draw=none]
      	box {$p_{\pi_1(\pi_0(1))}$} p[1];
      	[draw=none]
      	box {$p_{\pi_1(\pi_0(2))}$} p[2];
      	[draw=none]
      	box {$p_{\pi_1(\pi_0(3))}$} p[3];		
      	[name=g1]
		cnot p[3] | p[1];	
		box {$\pi_2\in\Pi$} (p);
		[draw=none]
      	box {$p_{\pi_2(\pi_1(\pi_0(0)))}$} p[0];
      	[draw=none]
      	box {$p_{\pi_2(\pi_1(\pi_0(1)))}$} p[1];
      	[draw=none]
      	box {$p_{\pi_2(\pi_1(\pi_0(2)))}$} p[2];
      	[draw=none]
      	box {$p_{\pi_2(\pi_1(\pi_0(3)))}$} p[3];
		[name=g2]
		cnot p[2] | p[0];	
		box {$\pi_3\in\Pi$} (p);
		[draw=none]
      	box {$p_{\pi_3(\pi_2(\pi_1(\pi_0(0))))}$} p[0];
      	[draw=none]
      	box {$p_{\pi_3(\pi_2(\pi_1(\pi_0(1))))}$} p[1];
      	[draw=none]
      	box {$p_{\pi_3(\pi_2(\pi_1(\pi_0(2))))}$} p[2];
      	[draw=none]
      	box {$p_{\pi_3(\pi_2(\pi_1(\pi_0(3))))}$} p[3];
		[name=g3]
		cnot p[1] | p[0];	
 	\end{yquant}

	\end{tikzpicture}
}
	\caption{Symbolic formulation for mapping the circuit shown in \autoref{fig:sample}}
	\vspace*{-1.5em}
	\label{fig:formulation}
\end{figure}
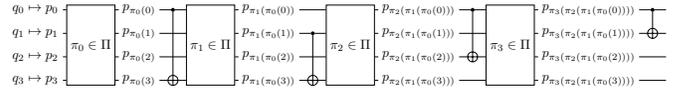

\begin{example}\label{ex:symbolic}
	Assume again that the circuit shown in \autoref{fig:sample} shall be mapped to the linear architecture shown in \autoref{fig:linear_architecture}. 
	Let $\Pi$ denote the set of all permutations of four elements.
	Then, \autoref{fig:formulation} sketches a symbolic formulation for the mapping task.
	Conceptually, any permutation $\pi\in\Pi$ can be applied in front of every gate of the circuit.
	This amounts to \mbox{$|G|*m!=4*4!=96$} permutations to be considered for determining an optimal solution.
\end{example}

Several solutions have been proposed for tackling the resulting complexity.
In~\cite{hirataEfficientConversionQuantum2011}, an exhaustive method is presented that are extended to heuristics for larger problems.
\emph{Siraichi et al.} used dynamic programming to determine an optimal solution~\cite{siraichiQubitAllocation2018}, while \emph{de Almeida et al.} formulated the mapping task as an integer linear programming problem.
\emph{Wille et al.} proposed a SAT formulation for the mapping problem in~\cite{willeMappingQuantumCircuits2019}.
A systematic enumeration and pruning technique is presented in~\cite{zhuExactQubitAllocation2020}.
Furthermore, there are works seeking a \emph{time-optimal} mapping using SAT~\cite{tanOptimalLayoutSynthesis2020} or guided search~\cite{zhangTimeoptimalQubitMapping2021}.

All these methods have in common that they eventually explore huge parts of the immense search space in order to determine an optimal solution. The question arises whether the search space can somehow be limited while preserving optimality.

\section{Limiting the Search Space}\label{sec:general}

As reviewed above, methods proposed thus far to solve the quantum circuit mapping problem extensively explore the search space spanned by considering \emph{all} possible permutations $\pi \in \Pi$ in front of every gate in order to guarantee an optimal solution.
As a result, the size of the search space significantly limits the efficiency of all those approaches.

In this work, we show that the search space of the mapping problem can be significantly limited, \emph{while still guaranteeing optimal results}. This is motivated by three observations which are described in this section. Based on that, we afterwards provide a theoretical argument confirming that these observations indeed preserve optimality (in Section~\ref{sec:formal}) and propose strategies how these findings can be utilized in the methods proposed before (in Section~\ref{sec:heuristics}). 

The first observation is based on the maximum length of all pairwise shortest paths between two nodes (i.e., physical qubits) in a coupling graph $(V, E)$, i.e., the longest, direct connection between two nodes. 
This length is~$K$ in the following and can be determined in $\mathcal{O}(|V|^3)$ using, e.g., the Floyd-Warshall algorithm~\cite{cormenIntroductionAlgorithmsThird2009}. 
We observe that it is sufficient to permute just enough so that any two qubits can interact with each other, i.e., instead of all $\pi\in\Pi$ permutations, it is sufficient to only consider permutations which can be realized by at most $K-1$ \textit{SWAP} operations (a more formal argumentation for that is presented later in \autoref{sec:theory}).
The more connected the coupling graph (i.e., the smaller $K$ in relation to $|V|$, or the larger $|E|$), the easier it is to reach all other qubits from any single qubit.
Consequently, this architecture-dependent limitation is most effective whenever the considered architecture is highly-connected.

\begin{example}\label{ex:perm_reduction}
	Consider the linear $4$-qubit architecture shown in \autoref{fig:linear_architecture}.
	Then, the longest (direct) connection involves $p_0$ and $p_3$---hence, \mbox{$K=3$}. Allowing only up to $K-1=2$ \textit{SWAP} operations per permutation reduces the number of permutations to be considered in front of every gate from $4!=24$ down to $9$, i.e., by more than half!
\end{example}

In addition to the above observation, another way to reduce the effective permutations that need to be considered when mapping a quantum circuit is already demonstrated in~\cite{willeMappingQuantumCircuits2019}.
Assume a quantum circuit uses less qubits than the architecture provides (i.e., $n < m$). 
Then, instead of considering the whole architecture at once, one can consider the mapping problem on each connected subgraph  of the targeted architecture composed of~$n$ nodes---leading to substantially smaller problems to solve. While there are at most $\binom{m}{n}$ potential subgraphs, the sparse nature of typical coupling graphs implies that the actual number of instances is much lower---consequently reducing the overall number of permutations. 
By limiting the search space for all the different problem instances on various subgraphs, the number of permutations is reduced even further.

\begin{example}\label{ex:subset-reduction}
	Assume the circuit shown in \autoref{fig:sample} shall be mapped to a $5$-qubit linear architecture. Instead of having to consider $5!*4 = 480$ permutations, there are only $4!*4 = 96$ per connected sub-graph of four vertices. Since there are only two such subgraphs in the linear architecture (that look exactly like the architecture shown in \autoref{fig:linear_architecture}), this amounts to a total of $192$ permutations to be considered, i.e., a reduction by \SI{60}{\percent}.
\end{example}

Finally, one can observe that the only SWAPs that are relevant before an operation, are those that change the position of either of the operation's qubits. All others can be delayed to a later point in time, as they do not serve to make the gate executable. Thus, any permutation that does not change the position of either of an operation's qubit may be ignored\footnote{ 
This observation assumes knowledge of the current mapping before each gate and is not feasible for blackbox implementations of the mapping problem such as SAT formulations, but rather applicable to iterative techniques such as informed search algorithms.}. 

\begin{example}
	Assume a \textit{CNOT} operation between $p_0$ and $p_3$ in a linear $4$-qubit architecture (as shown in \autoref{fig:linear_architecture}) shall be applied. Then, the permutations corresponding to the identity or swapping the middle qubits $p_1$ and $p_2$ can be ignored as they cannot possibly change the executability of the gate. In general, the larger the architecture, the more permutations can be ignored in front of every gate.
\end{example}

Overall, this shows substantial potential in limiting the search space of the problem and, by this, for improving the efficiency of corresponding optimal methods.  

\section{Preservation of Optimality\\When Limiting the Search Space}\label{sec:formal}

While the ideas presented above are merely based on observations, this section provides a formal argument why applying the observations summarized above still yields optimal results. To this end, we are going to use concepts from group theory~\cite{carterVisualGroupTheory2009}, which are briefly revisited first. 

\subsection{Group Theory and Permutation Groups}\label{sec:grouptheory}

A group $(G, *)$ is a set of elements $G$ equipped with a binary operation $*: G\times G \to G$ such that 
\begin{itemize}
	\item $\forall a, b, c\in G\colon (a * b) * c = a * (b * c)$ (\emph{associativity}),
	\item $\exists e \in G\;\forall g\in G\colon g * e = e * g = g$ (\emph{identity element}),
	\item $\forall g\in G \; \exists g'\in G\colon g * g' = g' * g = e$ (\emph{inverse element}).
\end{itemize}
If clear from the context, we will omit the $*$ when denoting group operations, i.e., we will write $g_0g_1$ instead of $g_0 * g_1$, and we will denote the inverse of an element $g\in G$ by $g^{-1}$.

\begin{example}
Consider the set $\Pi$ of all \emph{permutations} of a finite set of elements $P = \{p_0, \dots, p_{n-1}\}$.
Any permutation can be written as a set of cycles where cycles of length one are typically not denoted explicitly, e.g., $(01)(23)$ describes a permutation where $p_0 \leftrightarrow p_1$ and $p_2 \leftrightarrow p_3$, while all other elements remain unchanged.
 
\noindent The composition of two permutations $\pi_0$ and $\pi_1$ is defined as 
\begin{align*}
(\pi_1\circ\pi_0)\colon P &\to P \\
p &\mapsto \pi_1(\pi_0(p)).
\end{align*}
It can be shown that $(\Pi, \circ)$ forms a group---the so-called \emph{permutation group}.
Furthermore, it can easily be shown that the permutation group over a set of size $n$ consists of $n!$ elements.
\end{example}

An important concept in group theory is that of \emph{generating sets} of a group.
A subset of elements $S \subseteq G$ of a group  $(G, *)$ is called a generating set of the group if all elements from $G$ can be generated by (repeatedly) applying operation $*$ with the elements from $S$ to themselves or each other.

\begin{example}\label{ex:group_generate}
	Consider the group of permutations over the four element set \mbox{$P=\{p_0, p_1, p_2, p_3\}$}, which contains \mbox{$4! = 24$} elements. 
	Then, the set \mbox{$S = \{(01), (12), (23)\}$} consisting of three \mbox{nearest-neighbour} swaps already generates the whole group,~e.g., $(012) = (01) \circ (12)$.
\end{example}

The structure of a group $(G, *)$ with respect to a generating set $S\subseteq G$ can be represented as a directed graph---the \emph{Cayley graph}~\cite{carterVisualGroupTheory2009}. 
To this end, the graphs' vertices are given by the group elements $g \in G$ and, for every $s\in S$, there is an edge $g \xrightarrow{s} g'$ with $g,g'\in G$, if $s * g = g'$.

\begin{figure}[t]
\centering
\resizebox{0.85\linewidth}{!}{
\begin{tikzpicture}
\node (id) at (27.0bp,125.44bp) [draw,ellipse] {$()$};
  \node (p01) at (117.0bp,125.44bp) [draw,ellipse] {$(01)$};
  \node (p12) at (117.0bp,80.435bp) [draw,ellipse] {$(12)$};
  \node (p23) at (117.0bp,170.44bp) [draw,ellipse] {$(23)$};
  \node (p021) at (215.36bp,58.435bp) [draw,ellipse] {$(021)$};
  \node (p01_23) at (215.36bp,148.44bp) [draw,ellipse] {$(01)(23)$};
  \node (p012) at (215.36bp,13.435bp) [draw,ellipse] {$(012)$};
  \node (p132) at (215.36bp,103.44bp) [draw,ellipse] {$(132)$};
  \node (p123) at (215.36bp,193.44bp) [draw,ellipse] {$(123)$};
  \node (p02) at (317.82bp,13.435bp) [draw,ellipse] {$(02)$};
  \node (p0321) at (317.82bp,58.435bp) [draw,ellipse] {$(0321)$};
  \node (p0231) at (317.82bp,148.44bp) [draw,ellipse] {$(0231)$};
  \node (p0132) at (317.82bp,103.44bp) [draw,ellipse] {$(0132)$};
  \node (p13) at (317.82bp,193.44bp) [draw,ellipse] {$(13)$};
  \node (p0123) at (317.82bp,238.44bp) [draw,ellipse] {$(0123)$};
  \node (p023) at (420.29bp,148.44bp) [draw,ellipse] {$(023)$};
  \node (p031) at (420.29bp,58.435bp) [draw,ellipse] {$(031)$};
  \node (p032) at (420.29bp,13.435bp) [draw,ellipse] {$(032)$};
  \node (p02_13) at (420.29bp,103.44bp) [draw,ellipse] {$(02)(13)$};
  \node (p013) at (420.29bp,193.44bp) [draw,ellipse] {$(013)$};
  \node (p03) at (522.76bp,103.44bp) [draw,ellipse] {$(03)$};
  \node (p0312) at (522.76bp,58.435bp) [draw,ellipse] {$(0312)$};
  \node (p0213) at (522.76bp,148.44bp) [draw,ellipse] {$(0213)$};
  \node (p03_12) at (625.23bp,103.44bp) [draw,ellipse] {$(03)(12)$};
  \definecolor{strokecolor}{rgb}{0.26,0.39,0.85};
  \draw [strokecolor,line width = 1.5mm] (id) ..controls (65.642bp,125.44bp) and (78.715bp,125.44bp)  .. (p01);
  \definecolor{strokecolor}{rgb}{1.0,0.88,0.1};
  \draw [strokecolor,line width = 1.5mm] (id) ..controls (61.596bp,108.26bp) and (82.438bp,97.598bp)  .. (p12);
  \definecolor{strokecolor}{rgb}{0.66,0.66,0.66};
  \draw [strokecolor,line width = 1.5mm] (id) ..controls (61.596bp,142.62bp) and (82.438bp,153.27bp)  .. (p23);
  \definecolor{strokecolor}{rgb}{1.0,0.88,0.1};
  \draw [strokecolor,line width = 1.5mm] (p01) ..controls (145.71bp,105.5bp) and (163.92bp,92.523bp)  .. (180.0bp,81.435bp) .. controls (185.91bp,77.361bp) and (192.44bp,72.96bp)  .. (p021);
  \definecolor{strokecolor}{rgb}{0.66,0.66,0.66};
  \draw [strokecolor,line width = 1.5mm] (p01) ..controls (154.86bp,134.23bp) and (171.13bp,138.12bp)  .. (p01_23);
  \definecolor{strokecolor}{rgb}{0.26,0.39,0.85};
  \draw [strokecolor,line width = 1.5mm] (p12) ..controls (145.71bp,60.496bp) and (163.92bp,47.523bp)  .. (180.0bp,36.435bp) .. controls (185.91bp,32.361bp) and (192.44bp,27.96bp)  .. (p012);
  \definecolor{strokecolor}{rgb}{0.66,0.66,0.66};
  \draw [strokecolor,line width = 1.5mm] (p12) ..controls (156.75bp,89.686bp) and (175.93bp,94.262bp)  .. (p132);
  \definecolor{strokecolor}{rgb}{0.26,0.39,0.85};
  \draw [strokecolor,line width = 1.5mm] (p23) ..controls (154.75bp,162.04bp) and (170.85bp,158.37bp)  .. (p01_23);
  \definecolor{strokecolor}{rgb}{1.0,0.88,0.1};
  \draw [strokecolor,line width = 1.5mm] (p23) ..controls (156.75bp,179.69bp) and (175.93bp,184.26bp)  .. (p123);
  \definecolor{strokecolor}{rgb}{0.26,0.39,0.85};
  \draw [strokecolor,line width = 1.5mm] (p021) ..controls (253.82bp,41.655bp) and (279.41bp,30.191bp)  .. (p02);
  \definecolor{strokecolor}{rgb}{0.66,0.66,0.66};
  \draw [strokecolor,line width = 1.5mm] (p021) ..controls (255.98bp,58.435bp) and (272.51bp,58.435bp)  .. (p0321);
  \definecolor{strokecolor}{rgb}{1.0,0.88,0.1};
  \draw [strokecolor,line width = 1.5mm] (p01_23) ..controls (262.51bp,148.44bp) and (275.32bp,148.44bp)  .. (p0231);
  \definecolor{strokecolor}{rgb}{1.0,0.88,0.1};
  \draw [strokecolor,line width = 1.5mm] (p012) ..controls (257.37bp,13.435bp) and (275.96bp,13.435bp)  .. (p02);
  \definecolor{strokecolor}{rgb}{0.66,0.66,0.66};
  \draw [strokecolor,line width = 1.5mm] (p012) ..controls (239.25bp,27.226bp) and (245.5bp,31.705bp)  .. (250.71bp,36.435bp) .. controls (269.41bp,53.426bp) and (268.55bp,62.87bp)  .. (286.71bp,80.435bp) .. controls (290.97bp,84.551bp) and (295.95bp,88.615bp)  .. (p0132);
  \definecolor{strokecolor}{rgb}{0.26,0.39,0.85};
  \draw [strokecolor,line width = 1.5mm] (p132) ..controls (255.98bp,103.44bp) and (272.51bp,103.44bp)  .. (p0132);
  \definecolor{strokecolor}{rgb}{1.0,0.88,0.1};
  \draw [strokecolor,line width = 1.5mm] (p132) ..controls (239.25bp,117.23bp) and (245.5bp,121.7bp)  .. (250.71bp,126.44bp) .. controls (269.41bp,143.43bp) and (268.55bp,152.87bp)  .. (286.71bp,170.44bp) .. controls (291.16bp,174.73bp) and (296.4bp,178.98bp)  .. (p13);
  \definecolor{strokecolor}{rgb}{0.66,0.66,0.66};
  \draw [strokecolor,line width = 1.5mm] (p123) ..controls (257.37bp,193.44bp) and (275.96bp,193.44bp)  .. (p13);
  \definecolor{strokecolor}{rgb}{0.26,0.39,0.85};
  \draw [strokecolor,line width = 1.5mm] (p123) ..controls (253.36bp,210.01bp) and (278.09bp,221.09bp)  .. (p0123);
  \definecolor{strokecolor}{rgb}{0.26,0.39,0.85};
  \draw [strokecolor,line width = 1.5mm] (p0231) ..controls (363.07bp,148.44bp) and (379.7bp,148.44bp)  .. (p023);
  \definecolor{strokecolor}{rgb}{0.66,0.66,0.66};
  \draw [strokecolor,line width = 1.5mm] (p0231) ..controls (339.69bp,133.62bp) and (344.68bp,129.55bp)  .. (348.94bp,125.44bp) .. controls (367.1bp,107.87bp) and (366.23bp,98.426bp)  .. (384.94bp,81.435bp) .. controls (390.14bp,76.705bp) and (396.4bp,72.226bp)  .. (p031);
  \definecolor{strokecolor}{rgb}{0.66,0.66,0.66};
  \draw [strokecolor,line width = 1.5mm] (p02) ..controls (359.84bp,13.435bp) and (378.43bp,13.435bp)  .. (p032);
  \definecolor{strokecolor}{rgb}{1.0,0.88,0.1};
  \draw [strokecolor,line width = 1.5mm] (p0321) ..controls (363.07bp,58.435bp) and (379.7bp,58.435bp)  .. (p031);
  \definecolor{strokecolor}{rgb}{0.26,0.39,0.85};
  \draw [strokecolor,line width = 1.5mm] (p0321) ..controls (357.76bp,40.997bp) and (382.59bp,29.874bp)  .. (p032);
  \definecolor{strokecolor}{rgb}{1.0,0.88,0.1};
  \draw [strokecolor,line width = 1.5mm] (p0132) ..controls (360.18bp,103.44bp) and (372.99bp,103.44bp)  .. (p02_13);
  \definecolor{strokecolor}{rgb}{0.26,0.39,0.85};
  \draw [strokecolor,line width = 1.5mm] (p13) ..controls (359.84bp,193.44bp) and (378.43bp,193.44bp)  .. (p013);
  \definecolor{strokecolor}{rgb}{1.0,0.88,0.1};
  \draw [strokecolor,line width = 1.5mm] (p0123) ..controls (339.69bp,223.62bp) and (344.68bp,219.55bp)  .. (348.94bp,215.44bp) .. controls (367.1bp,197.87bp) and (366.23bp,188.43bp)  .. (384.94bp,171.44bp) .. controls (390.14bp,166.7bp) and (396.4bp,162.23bp)  .. (p023);
  \definecolor{strokecolor}{rgb}{0.66,0.66,0.66};
  \draw [strokecolor,line width = 1.5mm] (p0123) ..controls (357.76bp,221.0bp) and (382.59bp,209.87bp)  .. (p013);
  \definecolor{strokecolor}{rgb}{0.66,0.66,0.66};
  \draw [strokecolor,line width = 1.5mm] (p023) ..controls (458.75bp,131.66bp) and (484.35bp,120.19bp)  .. (p03);
  \definecolor{strokecolor}{rgb}{0.26,0.39,0.85};
  \draw [strokecolor,line width = 1.5mm] (p031) ..controls (458.75bp,75.215bp) and (484.35bp,86.679bp)  .. (p03);
  \definecolor{strokecolor}{rgb}{1.0,0.88,0.1};
  \draw [strokecolor,line width = 1.5mm] (p032) ..controls (458.3bp,30.01bp) and (483.03bp,41.086bp)  .. (p0312);
  \definecolor{strokecolor}{rgb}{0.66,0.66,0.66};
  \draw [strokecolor,line width = 1.5mm] (p02_13) ..controls (460.73bp,85.77bp) and (483.7bp,75.482bp)  .. (p0312);
  \definecolor{strokecolor}{rgb}{0.26,0.39,0.85};
  \draw [strokecolor,line width = 1.5mm] (p02_13) ..controls (460.73bp,121.1bp) and (483.7bp,131.39bp)  .. (p0213);
  \definecolor{strokecolor}{rgb}{1.0,0.88,0.1};
  \draw [strokecolor,line width = 1.5mm] (p013) ..controls (458.3bp,176.86bp) and (483.03bp,165.78bp)  .. (p0213);
  \definecolor{strokecolor}{rgb}{1.0,0.88,0.1};
  \draw [strokecolor,line width = 1.5mm] (p03) ..controls (561.99bp,103.44bp) and (576.52bp,103.44bp)  .. (p03_12);
  \definecolor{strokecolor}{rgb}{0.26,0.39,0.85};
  \draw [strokecolor,line width = 1.5mm] (p0312) ..controls (561.66bp,75.409bp) and (584.63bp,85.699bp)  .. (p03_12);
  \definecolor{strokecolor}{rgb}{0.66,0.66,0.66};
  \draw [strokecolor,line width = 1.5mm] (p0213) ..controls (561.66bp,131.46bp) and (584.63bp,121.17bp)  .. (p03_12);	
\end{tikzpicture}} 
  \caption{Cayley graph for the permutation group over $P=\{p_0, p_1, p_2, p_3\}$ using generators \textcolor{Blue01}{$(01)$}, \textcolor{Yellow12}{$(12)$}, \textcolor{Gray23}{$(23)$}.}
  \vspace*{-0.5em}
  \label{fig:perm-graph}
\end{figure}
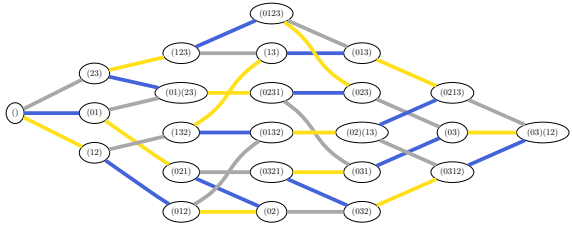

\begin{example}\label{ex:cayley}
	Consider again the permutation group over the set \mbox{$P=\{p_0, p_1, p_2, p_3\}$} and the set of generators \mbox{$S=\{$\textcolor{Blue01}{$(01)$}, \textcolor{Yellow12}{$(12)$}, \textcolor{Gray23}{$(23)$}$\}$}. Then, \autoref{fig:perm-graph} shows the corresponding Cayley graph.
\end{example}

\subsection{Theoretical Consideration}\label{sec:theory}

In order to understand why the total number of permutations can be reduced drastically while still preserving optimality, we look at the mapping problem from a group theoretic viewpoint.
Given a coupling graph $(V, E)$, the permutation group over~$|V|$ elements can be generated by the set of nearest-neighbour \textit{SWAP} operations executable on the coupling graph, i.e.,
\[S = \{(ij)\colon \forall e_{ij}\in E\}\]
 generates all possible permutations of $|V|$ elements for a particular architecture.
 
 \begin{example}
 	 The Cayley graph shown in \autoref{fig:perm-graph} illustrates that \mbox{$S=\{$\textcolor{Blue01}{$(01)$}, \textcolor{Yellow12}{$(12)$}, \textcolor{Gray23}{$(23)$}$\}$} generates the set of all permutations of four elements. Following any edge with the color corresponding to the generator shows the effect of applying it to the state denoted in the corresponding node. 
 	This specific set of generators corresponds to a linear $4$-qubit architecture given by \mbox{$(V, E) = (\{0,1,2,3\}, \{e_{01},e_{12},e_{23}\})$} as shown in \autoref{fig:linear_architecture}.
 \end{example}
 
Now, let $K$ be the maximum length of all pairwise shortest paths between two nodes, i.e., the longest, direct connection between any two nodes in the coupling graph.
Then, the reduced set of permutations (denoted $\Pi'$ in the following) is composed of all permutations that can be generated by applying at most $K-1$ generators from $S$.
This is understood as constructing the Cayley graph associated to $S$, starting at the identity, and stopping after $K-1$ steps. 
 
 \begin{figure}[t]
\centering
\resizebox{0.85\linewidth}{!}{
\begin{tikzpicture}
\node (id) at (27.0bp,125.44bp) [draw,ellipse] {$()$};
  \node (p01) at (117.0bp,125.44bp) [draw,ellipse] {$(01)$};
  \node (p12) at (117.0bp,80.435bp) [draw,ellipse] {$(12)$};
  \node (p23) at (117.0bp,170.44bp) [draw,ellipse] {$(23)$};
  \node (p021) at (215.36bp,58.435bp) [draw,ellipse] {$(021)$};
  \node (p01_23) at (215.36bp,148.44bp) [draw,ellipse] {$(01)(23)$};
  \node (p012) at (215.36bp,13.435bp) [draw,ellipse] {$(012)$};
  \node (p132) at (215.36bp,103.44bp) [draw,ellipse] {$(132)$};
  \node (p123) at (215.36bp,193.44bp) [draw,ellipse] {$(123)$};
  
  \node (p02) at (317.82bp,13.435bp) [draw,ellipse,fill=gray!20] {$(02)$};
  \node (p0321) at (317.82bp,58.435bp) [draw,ellipse,fill=gray!20] {$(0321)$};
  \node (p0231) at (317.82bp,148.44bp) [draw,ellipse,fill=gray!20] {$(0231)$};
  \node (p0132) at (317.82bp,103.44bp) [draw,ellipse,fill=gray!20] {$(0132)$};
  \node (p13) at (317.82bp,193.44bp) [draw,ellipse,fill=gray!20] {$(13)$};
  \node (p0123) at (317.82bp,238.44bp) [draw,ellipse,fill=gray!20] {$(0123)$};
  \node (p023) at (420.29bp,148.44bp) [draw,ellipse,fill=gray!20] {$(023)$};
  \node (p031) at (420.29bp,58.435bp) [draw,ellipse,fill=gray!20] {$(031)$};
  \node (p032) at (420.29bp,13.435bp) [draw,ellipse,fill=gray!20] {$(032)$};
  \node (p02_13) at (420.29bp,103.44bp) [draw,ellipse,fill=gray!20] {$(02)(13)$};
  \node (p013) at (420.29bp,193.44bp) [draw,ellipse,fill=gray!20] {$(013)$};
  \node (p03) at (522.76bp,103.44bp) [draw,ellipse,fill=gray!20] {$(03)$};
  \node (p0312) at (522.76bp,58.435bp) [draw,ellipse,fill=gray!20] {$(0312)$};
  \node (p0213) at (522.76bp,148.44bp) [draw,ellipse,fill=gray!20] {$(0213)$};
  \node (p03_12) at (625.23bp,103.44bp) [draw,ellipse,fill=gray!20] {$(03)(12)$};
  \definecolor{strokecolor}{rgb}{0.26,0.39,0.85};
  \draw [strokecolor,line width = 1.5mm] (id) ..controls (65.642bp,125.44bp) and (78.715bp,125.44bp)  .. (p01);
  \definecolor{strokecolor}{rgb}{1.0,0.88,0.1};
  \draw [strokecolor,line width = 1.5mm] (id) ..controls (61.596bp,108.26bp) and (82.438bp,97.598bp)  .. (p12);
  \definecolor{strokecolor}{rgb}{0.66,0.66,0.66};
  \draw [strokecolor,line width = 1.5mm] (id) ..controls (61.596bp,142.62bp) and (82.438bp,153.27bp)  .. (p23);
  \definecolor{strokecolor}{rgb}{1.0,0.88,0.1};
  \draw [strokecolor,line width = 1.5mm] (p01) ..controls (145.71bp,105.5bp) and (163.92bp,92.523bp)  .. (180.0bp,81.435bp) .. controls (185.91bp,77.361bp) and (192.44bp,72.96bp)  .. (p021);
  \definecolor{strokecolor}{rgb}{0.66,0.66,0.66};
  \draw [strokecolor,line width = 1.5mm] (p01) ..controls (154.86bp,134.23bp) and (171.13bp,138.12bp)  .. (p01_23);
  \definecolor{strokecolor}{rgb}{0.26,0.39,0.85};
  \draw [strokecolor,line width = 1.5mm] (p12) ..controls (145.71bp,60.496bp) and (163.92bp,47.523bp)  .. (180.0bp,36.435bp) .. controls (185.91bp,32.361bp) and (192.44bp,27.96bp)  .. (p012);
  \definecolor{strokecolor}{rgb}{0.66,0.66,0.66};
  \draw [strokecolor,line width = 1.5mm] (p12) ..controls (156.75bp,89.686bp) and (175.93bp,94.262bp)  .. (p132);
  \definecolor{strokecolor}{rgb}{0.26,0.39,0.85};
  \draw [strokecolor,line width = 1.5mm] (p23) ..controls (154.75bp,162.04bp) and (170.85bp,158.37bp)  .. (p01_23);
  \definecolor{strokecolor}{rgb}{1.0,0.88,0.1};
  \draw [strokecolor,line width = 1.5mm] (p23) ..controls (156.75bp,179.69bp) and (175.93bp,184.26bp)  .. (p123);
  \definecolor{strokecolor}{rgb}{0.26,0.39,0.85};
  \draw [strokecolor!20,line width = 1.5mm] (p021) ..controls (253.82bp,41.655bp) and (279.41bp,30.191bp)  .. (p02);
  \definecolor{strokecolor}{rgb}{0.66,0.66,0.66};
  \draw [strokecolor!20,line width = 1.5mm] (p021) ..controls (255.98bp,58.435bp) and (272.51bp,58.435bp)  .. (p0321);
  \definecolor{strokecolor!20}{rgb}{1.0,0.88,0.1};
  \draw [strokecolor!20,line width = 1.5mm] (p01_23) ..controls (262.51bp,148.44bp) and (275.32bp,148.44bp)  .. (p0231);
  \definecolor{strokecolor!20}{rgb}{1.0,0.88,0.1};
  \draw [strokecolor!20,line width = 1.5mm] (p012) ..controls (257.37bp,13.435bp) and (275.96bp,13.435bp)  .. (p02);
  \definecolor{strokecolor!20}{rgb}{0.66,0.66,0.66};
  \draw [strokecolor!20,line width = 1.5mm] (p012) ..controls (239.25bp,27.226bp) and (245.5bp,31.705bp)  .. (250.71bp,36.435bp) .. controls (269.41bp,53.426bp) and (268.55bp,62.87bp)  .. (286.71bp,80.435bp) .. controls (290.97bp,84.551bp) and (295.95bp,88.615bp)  .. (p0132);
  \definecolor{strokecolor}{rgb}{0.26,0.39,0.85};
  \draw [strokecolor!20,line width = 1.5mm] (p132) ..controls (255.98bp,103.44bp) and (272.51bp,103.44bp)  .. (p0132);
  \definecolor{strokecolor}{rgb}{1.0,0.88,0.1};
  \draw [strokecolor!20,line width = 1.5mm] (p132) ..controls (239.25bp,117.23bp) and (245.5bp,121.7bp)  .. (250.71bp,126.44bp) .. controls (269.41bp,143.43bp) and (268.55bp,152.87bp)  .. (286.71bp,170.44bp) .. controls (291.16bp,174.73bp) and (296.4bp,178.98bp)  .. (p13);
  \definecolor{strokecolor}{rgb}{0.66,0.66,0.66};
  \draw [strokecolor!20,line width = 1.5mm] (p123) ..controls (257.37bp,193.44bp) and (275.96bp,193.44bp)  .. (p13);
  \definecolor{strokecolor}{rgb}{0.26,0.39,0.85};
  \draw [strokecolor!20,line width = 1.5mm] (p123) ..controls (253.36bp,210.01bp) and (278.09bp,221.09bp)  .. (p0123);
  \definecolor{strokecolor}{rgb}{0.26,0.39,0.85};
  \draw [strokecolor!20,line width = 1.5mm] (p0231) ..controls (363.07bp,148.44bp) and (379.7bp,148.44bp)  .. (p023);
  \definecolor{strokecolor}{rgb}{0.66,0.66,0.66};
  \draw [strokecolor!20,line width = 1.5mm] (p0231) ..controls (339.69bp,133.62bp) and (344.68bp,129.55bp)  .. (348.94bp,125.44bp) .. controls (367.1bp,107.87bp) and (366.23bp,98.426bp)  .. (384.94bp,81.435bp) .. controls (390.14bp,76.705bp) and (396.4bp,72.226bp)  .. (p031);
  \definecolor{strokecolor}{rgb}{0.66,0.66,0.66};
  \draw [strokecolor!20,line width = 1.5mm] (p02) ..controls (359.84bp,13.435bp) and (378.43bp,13.435bp)  .. (p032);
  \definecolor{strokecolor}{rgb}{1.0,0.88,0.1};
  \draw [strokecolor!20,line width = 1.5mm] (p0321) ..controls (363.07bp,58.435bp) and (379.7bp,58.435bp)  .. (p031);
  \definecolor{strokecolor}{rgb}{0.26,0.39,0.85};
  \draw [strokecolor!20,line width = 1.5mm] (p0321) ..controls (357.76bp,40.997bp) and (382.59bp,29.874bp)  .. (p032);
  \definecolor{strokecolor}{rgb}{1.0,0.88,0.1};
  \draw [strokecolor!20,line width = 1.5mm] (p0132) ..controls (360.18bp,103.44bp) and (372.99bp,103.44bp)  .. (p02_13);
  \definecolor{strokecolor}{rgb}{0.26,0.39,0.85};
  \draw [strokecolor!20,line width = 1.5mm] (p13) ..controls (359.84bp,193.44bp) and (378.43bp,193.44bp)  .. (p013);
  \definecolor{strokecolor}{rgb}{1.0,0.88,0.1};
  \draw [strokecolor!20,line width = 1.5mm] (p0123) ..controls (339.69bp,223.62bp) and (344.68bp,219.55bp)  .. (348.94bp,215.44bp) .. controls (367.1bp,197.87bp) and (366.23bp,188.43bp)  .. (384.94bp,171.44bp) .. controls (390.14bp,166.7bp) and (396.4bp,162.23bp)  .. (p023);
  \definecolor{strokecolor}{rgb}{0.66,0.66,0.66};
  \draw [strokecolor!20,line width = 1.5mm] (p0123) ..controls (357.76bp,221.0bp) and (382.59bp,209.87bp)  .. (p013);
  \definecolor{strokecolor}{rgb}{0.66,0.66,0.66};
  \draw [strokecolor!20,line width = 1.5mm] (p023) ..controls (458.75bp,131.66bp) and (484.35bp,120.19bp)  .. (p03);
  \definecolor{strokecolor}{rgb}{0.26,0.39,0.85};
  \draw [strokecolor!20,line width = 1.5mm] (p031) ..controls (458.75bp,75.215bp) and (484.35bp,86.679bp)  .. (p03);
  \definecolor{strokecolor}{rgb}{1.0,0.88,0.1};
  \draw [strokecolor!20,line width = 1.5mm] (p032) ..controls (458.3bp,30.01bp) and (483.03bp,41.086bp)  .. (p0312);
  \definecolor{strokecolor}{rgb}{0.66,0.66,0.66};
  \draw [strokecolor!20,line width = 1.5mm] (p02_13) ..controls (460.73bp,85.77bp) and (483.7bp,75.482bp)  .. (p0312);
  \definecolor{strokecolor}{rgb}{0.26,0.39,0.85};
  \draw [strokecolor!20,line width = 1.5mm] (p02_13) ..controls (460.73bp,121.1bp) and (483.7bp,131.39bp)  .. (p0213);
  \definecolor{strokecolor}{rgb}{1.0,0.88,0.1};
  \draw [strokecolor!20,line width = 1.5mm] (p013) ..controls (458.3bp,176.86bp) and (483.03bp,165.78bp)  .. (p0213);
  \definecolor{strokecolor}{rgb}{1.0,0.88,0.1};
  \draw [strokecolor!20,line width = 1.5mm] (p03) ..controls (561.99bp,103.44bp) and (576.52bp,103.44bp)  .. (p03_12);
  \definecolor{strokecolor}{rgb}{0.26,0.39,0.85};
  \draw [strokecolor!20,line width = 1.5mm] (p0312) ..controls (561.66bp,75.409bp) and (584.63bp,85.699bp)  .. (p03_12);
  \definecolor{strokecolor}{rgb}{0.66,0.66,0.66};
  \draw [strokecolor!20,line width = 1.5mm] (p0213) ..controls (561.66bp,131.46bp) and (584.63bp,121.17bp)  .. (p03_12);	
\end{tikzpicture}} 
  \caption{Reduced Cayley graph for the linear $4$-qubit architecture from \autoref{fig:linear_architecture}}
  \label{fig:perm-graph-reduced}
\end{figure}
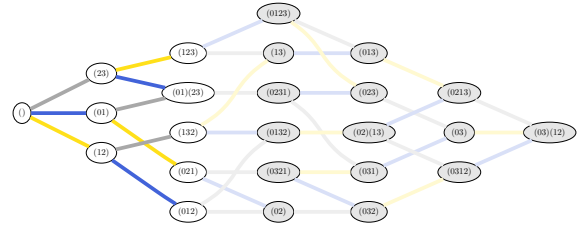
 
 \begin{example}
 	Consider again the linear $4$-qubit architecture shown in \autoref{fig:linear_architecture} with \mbox{$S=\{$\textcolor{Blue01}{$(01)$}, \textcolor{Yellow12}{$(12)$}, \textcolor{Gray23}{$(23)$}$\}$}. Then, \autoref{fig:perm-graph-reduced} shows the reduced Cayley graph for this architecture. As already seen in \autoref{ex:perm_reduction}, only $9$ permutations need to be considered here (instead of $24$). 
 \end{example} 
 
\begin{figure}[t]
    \centering
 	\resizebox{0.9\linewidth}{!}{
	 \begin{tikzpicture}
	\begin{yquant}
		nobit q[4];
		
		[frameless]
		subcircuit {
			[out]
			qubit {$\ket{p_{i}}$} q[1];
			[out]
			qubit {$\ket{p_{j}}$} q[+1];
			[out]
			qubit {$\ket{p_{k}}$} q[+1];
			[out]
			qubit {$\ket{p_{l}}$} q[+1];
			[this subcircuit box style={dashed, "$\pi\in\Pi\setminus\Pi'$"}]
			subcircuit {
			qubit {} q[4];
	    		box {$\pi'\in\Pi'$} (q);
	    		swap (q[2],q[3]);
	    		} (q);
	    		align q[1], q[2];
	    		cnot q[1] | q[0];
	    		box {$\pi'\in\Pi'$} (q);
	    		cnot q[2] | q[1];
		} (q);
		discard -;
		
		[draw=none]
		box {$=$} (q);
		
		[frameless]
		subcircuit {
			[out]
			qubit {$\ket{p_{i}}$} q[1];
			[out]
			qubit {$\ket{p_{j}}$} q[+1];
			[out]
			qubit {$\ket{p_{k}}$} q[+1];
			[out]
			qubit {$\ket{p_{l}}$} q[+1];
	    		box {$\pi'\in\Pi'$} (q);
			cnot q[1] | q[0];
	    		align q[1], q[2];
    		box {$\pi'\in\Pi'$} (q);
	    		cnot q[2] | q[1];
		} (q);	
		discard -;
	\end{yquant}
\end{tikzpicture}}
    \hfill    
    \caption{Main circuit identity that allows to show optimality is preserved}\label{fig:executionorder}\vspace*{-0em}
\end{figure}
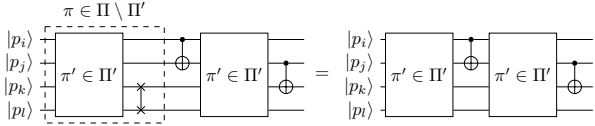

 In order to show that only considering the reduced permutation set $\Pi'$ preserves optimality, assume otherwise, i.e., assume that at any point in the mapping there exists a permutation $\pi\in\Pi\setminus \Pi'$ that allows for a cheaper overall mapping.
 More specifically, assume that in front of a $CNOT(q_c, q_t)$ (where $q_c$ is currently mapped to $p_i$ and $q_t$ to $p_j$) there exists $\pi \in \Pi \setminus \Pi'$ such that 
 \begin{enumerate}
 	\item the gate remains executable (i.e., satisfies the coupling constraints) and
 	\item the overall amount of \textit{SWAPs} needed to map the circuit is reduced.
 \end{enumerate}
 Since, $\pi\not\in\Pi'$, realizing $\pi$ must at least require $K$ \textit{SWAPs}.
 Assume that, without loss of generality,  $\pi = (kl)\circ \pi'$ for some $\pi'\in\Pi'$.
 Since $K$ denotes the longest, direct path between any two qubits, 
 realizing the operation on any other edge of the coupling graph could have already been done using at most $K-1$ \textit{SWAPs}.
 Due to $1)$, any \textit{SWAP} involving $i$ or $j$ that makes the gate non-executable is ruled out.
 Thus, $k \neq i, j \wedge l \neq i, j$.
 However, in that case 
 \[\resizebox{0.99\linewidth}{!}{$\mathit{SWAP}(p_k,p_l) \mathit{CNOT}(p_i,p_j) = \mathit{CNOT}(p_i,p_j) \mathit{SWAP}(p_k,p_l)$}\]
 holds, since gates acting on distinct sets of qubits commute. 
 Consequently, applying the \textit{SWAP} $(kl)$ in front of the current gate cannot reduce the overall cost of the resulting circuit, since $(kl)$ will be considered once a gate involving either qubit $k$ or $l$ is encountered later on---contradicting $2)$.
 \autoref{fig:executionorder} illustrates this central circuit identity.
 
 This is a strong argument that shows that optimality is preserved if the first observation presented in Section~\ref{sec:general} is applied. 
 A similar argument can be made to argue that the method of ignoring permutations which do not alter the qubits involved in an operation preserves optimality. 
While it is not per-se clear whether distributing the problem by considering all possible connected subgraphs preserves optimality, our experimental evaluations (which are summarized in \autoref{sec:results}) suggest this to be the case.

\section{Resulting Strategies}\label{sec:heuristics}

Based on the considerations from above, we are now proposing two strategies for reducing the number of permutations during quantum circuit mapping. 
As experiments (summarized in Section~\ref{sec:results}) confirm, they allow to substantially reduce the complexity and, hence, the run-time of corresponding optimal mapping methods.

\subsection{Architecture Limit}\label{sec:ideaA}
In order to capitalize on the first observation from \autoref{sec:general}, the length $K$ of the longest, direct path through the complete architecture has to be determined.
It is sufficient to compute this quantity once for a given architecture, e.g., by using the Floyd-Warshall algorithm~\cite{cormenIntroductionAlgorithmsThird2009}, and storing it for future reuse.

\begin{example}\label{ex:floyd}
	Computing all pairwise shortest paths for the linear $4$-qubit architecture shown in \autoref{fig:linear_architecture} results in the following tableau
	\[
	\begin{bsmallmatrix}
		0 & 1 & 2 & 3 \\
		 1 & 0 & 1 & 2 \\
		 2 & 1 & 0 & 1\\
		 3 & 2 & 1 & 0		
	\end{bsmallmatrix},
	\]
	which allows to determine $K=3$.
\end{example}

Once $K$ is calculated, the reduced set of permutations $\Pi'$ to be considered in front of every gate needs to be determined. 
Since this reduction only depends on the targeted architecture and is independent of the actual gates to be executed, it can also be computed once, e.g., by constructing a representation of the Cayley graph for the given architecture and stopping after $K-1$ applications.
By defining an ordering of all permutations (e.g., lexicographic ordering), a bitset of size $m!$ may be used to keep track of which permutations are enabled and which are not.

\begin{example}
	Computing the reduced Cayley graph for the linear $4$-qubit architecture results in
	\[\resizebox{0.95\linewidth}{!}{$\Pi'=\{(), (01), (12), (23), (123), (01)(23), (132), (021), (012)\},$}\]
	as previously shown in \autoref{fig:perm-graph-reduced}.
	Assuming lexicographic ordering, $\Pi'$ corresponds to $0000\; 0000\; 0001\; 0001\; 1101\; 1111$.
\end{example}

Using $\Pi'$ instead of $\Pi$ allows for optimal mapping of circuits considering a substantially smaller search space.
The strategy works best for quantum circuits having as many, or close to as many, qubits as the architecture they get mapped to. This is because this strategy capitalizes on the restrictions imposed due to the inherent structure of the given architecture and limits the number of permutations based on this. 

\subsection{Subgraph Limit}\label{sec:ideaB}
The second strategy combines dividing the problem into smaller problems on all connected subgraphs with the idea of limiting the number of \textit{SWAP} operations maximally considered in front of each gate\footnote{Whenever $n=m$, this naturally becomes the strategy from \autoref{sec:ideaA}.}.
In this fashion, the number of permutations to be considered is further reduced.
However, the length $K$ of the longest path now has to be calculated for every possible subgraph of the targeted architecture, but not the architecture itself.
Due to symmetry/regularity of many quantum architectures, many of these computations are redundant and can be skipped.
Afterwards, the subset $\Pi'\subseteq \Pi$ can be determined for every possible connected subgraph the same way as in \autoref{sec:ideaA}.

\begin{example}
	Assume the circuit shown in \autoref{fig:sample} is to be mapped to a linear $5$-qubit architecture. As shown in \autoref{ex:subset-reduction}, there are two $4$-qubit subsets of this architecture---both of which have precisely the structure shown in \autoref{fig:linear_architecture}.
	Consequently, only a single longest, direct path calculation has to be carried out, which results in the same tableau as in \autoref{ex:floyd}---and, hence, $K=3$.
	The corresponding subsets~$\Pi'$ are characterized by $0000\; 0000\; 0001\; 0001\; 1101\; 1111$ in both cases.
\end{example}
 
In general, the strategy's benefits are biggest whenever $n<<m$, i.e., whenever a circuit is mapped to a significantly larger architecture. 
The architecture itself also plays a vital role in the efficiency of this technique.
On the one hand, the less connected the architecture is, the fewer connected subsets there are to consider. 
On the other hand, the less connected the architecture, the larger $K$ (and the larger $K$, the more permutations have to be considered).
Consequently, there certainly exists a ``sweet spot'' of architecture connectivity for employing this strategy. 
As our experimental evaluations (which are summarized next) show, using this strategy leads to improvements in almost all considered cases, both compared to not using subgraph-division at all, as well as just using subgraph-division.

Since we have no proof that the strategy of considering all connected subgraphs preserves optimality, there is no guarantee that results obtained from the strategy proposed in this section remain optimal. However, our experimental evaluations indicate that optimality is preserved at least for all the benchmarks we ran our experiments on.
This shows great promise for applying optimal techniques for the mapping problem to larger circuits and/or larger architectures.

\section{Experimental Results}
\label{sec:results}

\begin{table*}[t]
	\sisetup{table-format=4.2, group-minimum-digits=4, table-number-alignment=right,table-text-alignment=right}
	\caption{Experimental Evaluations}
	\label{tbl:evaluation_results} 	
	\centering
	\resizebox{0.99\linewidth}{!}{
		\begin{tabular}{lrrr !{\qquad} rr !{\qquad} rrr !{\qquad} rr!{\qquad} rrr}\toprule
			\multicolumn{4}{c}{} & \multicolumn{5}{c}{Without Subgraphs} & \multicolumn{5}{c}{With Subgraphs} \\
			\cmidrule(l{4em}r{4em}){5-9}\cmidrule(l{4em}r{4em}){10-14}
			\multicolumn{4}{c}{Benchmark} & \multicolumn{2}{c}{JKQ QMAP~\cite{willeMappingQuantumCircuits2019}} & \multicolumn{3}{c}{Architecture Limit (\autoref{sec:ideaA})} & \multicolumn{2}{c}{JKQ QMAP~\cite{willeMappingQuantumCircuits2019}} & \multicolumn{3}{c}{Subgraph Limit (\autoref{sec:ideaB})} \\
			\cmidrule(l{2em}r{2em}){1-4}\cmidrule(lr{1em}){5-6}\cmidrule(lr{2em}){7-9}\cmidrule(l{1em}r{1em}){10-11}\cmidrule(l{1em}r{1em}){12-14}
			Name & {$n$} & {$|G|$} & {$c$} & {$|\Pi|$} & {$t_\mathit{ref}$~[\si{\second}]} & {$|\Pi'|$} & {$t_\mathit{prop}$~[\si{\second}]} & {$t_\mathit{ref}/t_\mathit{prop}$} & {$|\Pi|$} & {$t_\mathit{ref}$~[\si{\second}]} & {$|\Pi'|$} & {$t_\mathit{prop}$~[\si{\second}]} & {$t_\mathit{ref}/t_\mathit{prop}$}\\\midrule
			\begin{filecontents*}{results.csv}
4\_49\_16;217;5;London;5;26040;3600.0;207.0;171.15;3472.0;21.03;7.5;3600.0;-1.0;171.07;3472;21.04405298;7.5;26040.00
hwb4\_49;233;5;London;5;27960;3600.0;213.0;198.45;3728.0;18.14;7.5;3600.0;-1.0;199.57;3728;18.0386437;7.5;27960.00
mod10\_171;244;5;London;5;29280;3600.0;285.0;291.04;3904.0;12.37;7.5;3600.0;-1.0;283.86;3904;12.68214111;7.5;29280.00
mini-alu\_167;288;5;London;5;34560;3600.0;330.0;477.25;4608.0;7.54;7.5;3600.0;-1.0;475.06;4608;7.578003354;7.5;34560.00
one-two-three-v0\_97;290;5;London;5;34800;3600.0;234.0;364.29;4640.0;9.88;7.5;3600.0;-1.0;363.97;4640;9.890998607;7.5;34800.00
alu-v2\_31;451;5;London;5;54120;3600.0;375.0;1418.29;7216.0;2.54;7.5;3600.0;-1.0;1413.01;7216;2.547756715;7.5;54120.00
decod24-v3\_45;150;5;London;5;18000;1750.86;156.0;57.53;2400.0;30.43;7.5;1738.25;156.0;58.65;2400;29.63608129;7.5;18000.00
aj-e11\_165;151;5;London;5;18120;1691.31;129.0;47.9;2416.0;35.31;7.5;1689.52;129.0;47.94;2416;35.24509905;7.5;18120.00
4mod7-v1\_96;164;5;London;5;19680;1679.76;123.0;51.87;2624.0;32.39;7.5;1655.91;123.0;51.49;2624;32.16017514;7.5;19680.00
alu-v2\_32;163;5;London;5;19560;1481.97;117.0;48.19;2608.0;30.75;7.5;1477.13;117.0;47.59;2608;31.03587401;7.5;19560.00
4gt10-v1\_81;148;5;London;5;17760;1319.27;111.0;40.84;2368.0;32.3;7.5;1347.52;111.0;41.16;2368;32.74117369;7.5;17760.00
one-two-three-v0\_98;146;5;London;5;17520;1227.19;108.0;44.85;2336.0;27.36;7.5;1254.3;108.0;46.91;2336;26.73979801;7.5;17520.00
one-two-three-v1\_99;132;5;London;5;15840;1071.84;108.0;35.53;2112.0;30.16;7.5;1056.8;108.0;35.24;2112;29.98559136;7.5;15840.00
4gt5\_77;131;5;London;5;15720;914.56;99.0;31.16;2096.0;29.35;7.5;927.59;99.0;31.53;2096;29.4198008;7.5;15720.00
4gt13\_91;103;5;London;5;12360;632.21;93.0;20.17;1648.0;31.34;7.5;651.83;93.0;20.27;1648;32.14989614;7.5;12360.00
miller\_11;50;3;London;5;6000;624.63;27.0;0.33;800.0;1880.11;7.5;0.11;27.0;0.06;150;1.913175995;2;300.00
alu-v4\_36;115;5;London;5;13800;594.61;87.0;20.36;1840.0;29.2;7.5;586.32;87.0;20.7;1840;28.32644047;7.5;13800.00
4gt5\_76;91;5;London;5;10920;444.35;84.0;0.35;1456.0;1254.61;7.5;458.9;84.0;0.35;1456;1298.817972;7.5;10920.00
decod24-v1\_41;85;5;London;5;10200;344.17;84.0;11.1;1360.0;31.0;7.5;341.9;84.0;11.46;1360;29.83945419;7.5;10200.00
decod24-v2\_43;52;4;London;5;6240;254.36;27.0;0.24;832.0;1047.78;7.5;26.65;33.0;0.32;468;83.75106424;2.666666667;1248.00
4mod5-v1\_23;69;5;London;5;8280;198.41;66.0;6.55;1104.0;30.31;7.5;210.68;66.0;6.56;1104;32.11497013;7.5;8280.00
4gt13\_92;66;5;London;5;7920;193.88;78.0;5.79;1056.0;33.49;7.5;196.83;78.0;5.75;1056;34.23369204;7.5;7920.00
rd32\_270;84;5;London;5;10080;155.96;54.0;5.59;1344.0;27.89;7.5;155.42;54.0;5.64;1344;27.57701933;7.5;10080.00
4mod5-v0\_18;69;5;London;5;8280;135.65;48.0;3.52;1104.0;38.57;7.5;144.92;48.0;3.44;1104;42.14972402;7.5;8280.00
one-two-three-v2\_100;69;5;London;5;8280;133.85;48.0;0.23;1104.0;577.65;7.5;131.56;48.0;0.23;1104;580.9203934;7.5;8280.00
mod5d2\_64;53;5;London;5;6360;84.36;42.0;0.22;848.0;385.96;7.5;83.82;42.0;0.22;848;381.3554398;7.5;6360.00
alu-v1\_28;37;5;London;5;4440;65.02;30.0;0.13;592.0;506.02;7.5;65.82;30.0;0.13;592;508.1134696;7.5;4440.00
3\_17\_13;36;3;London;5;4320;35.31;18.0;0.26;576.0;137.53;7.5;0.08;18.0;0.04;108;1.917948749;2;216.00
rd32-v1\_68;36;4;London;5;4320;12.85;18.0;0.13;576.0;102.67;7.5;0.51;18.0;0.17;324;2.961089018;2.666666667;864.00
rd32-v0\_66;34;4;London;5;4080;12.83;18.0;0.12;544.0;103.1;7.5;0.51;18.0;0.17;306;2.969737618;2.666666667;816.00
4gt13\_90;107;5;London;5;12840;10.73;114.0;0.3;1712.0;35.36;7.5;10.9;114.0;0.3;1712;36.01626696;7.5;12840.00
mod5mils\_65;35;5;London;5;4200;2.53;24.0;0.12;560.0;21.51;7.5;2.49;24.0;0.12;560;21.12668808;7.5;4200.00
alu-v4\_37;37;5;London;5;4440;2.26;30.0;0.15;592.0;14.71;7.5;2.25;30.0;0.16;592;14.46772152;7.5;4440.00
one-two-three-v3\_101;70;5;London;5;8400;1.69;66.0;0.26;1120.0;6.45;7.5;1.65;66.0;0.27;1120;6.203257483;7.5;8400.00
alu-v3\_34;52;5;London;5;6240;1.57;51.0;0.19;832.0;8.44;7.5;1.6;51.0;0.19;832;8.599417198;7.5;6240.00
decod24-v0\_38;51;4;London;5;6120;1.33;27.0;0.31;816.0;4.26;7.5;35.77;36.0;0.82;459;43.49107617;2.666666667;1224.00
qe\_qft\_5;107;5;London;5;12840;0.99;9.0;0.22;1712.0;4.54;7.5;0.97;9.0;0.22;1712;4.414038603;7.5;12840.00
4mod5-v0\_19;35;5;London;5;4200;0.96;30.0;0.83;560.0;1.16;7.5;1.43;30.0;0.82;560;1.737189255;7.5;4200.00
4gt11\_82;27;5;London;5;3240;0.93;45.0;0.24;432.0;3.92;7.5;1.37;45.0;0.24;432;5.792336236;7.5;3240.00
alu-v3\_35;37;5;London;5;4440;0.87;30.0;0.15;592.0;5.63;7.5;1.07;30.0;0.15;592;6.931969165;7.5;4440.00
\end{filecontents*}
\csvreader[column count=19, no head, separator=semicolon, late after line=\\, late after last line=\\\bottomrule	]{results.csv}	
			{1=\Name, 2=\Layers, 3=\logQB, 4=\arch, 5=\pysQB, 6=\perm, 7=\baselineT, 8=\baselineG, 9=\architectT, 10=\architectPi, 11=\darchT, 12=\darchPi, 13=\subsT, 14=\subsG, 15=\subslT, 16=\subslPi, 17=\dsubslT, 18=\dsubslPi, 19=\subsPi}
			{%
			\Name  &
			\tablenum[table-format=1,zero-decimal-to-integer]{\logQB} & 
			\tablenum[table-format=3,zero-decimal-to-integer]{\Layers} &
			\ifthenelse{\lengthtest{\baselineG pt < 0 pt}}{-}{\tablenum[table-format=3,zero-decimal-to-integer]{\baselineG}} &
			\tablenum[table-format=5,zero-decimal-to-integer]{\perm} &
			\ifthenelse{\lengthtest{\baselineT pt > 3599 pt}}{\SI{>1}{\hour}}{\ifthenelse{\lengthtest{\baselineT pt < 0 pt}}{[MemOut]}{\tablenum[round-mode=places, round-precision=2]{\baselineT}}} &
			\tablenum[table-format=4,zero-decimal-to-integer]{\architectPi} &
			{\bfseries \tablenum[round-mode=places, round-precision=2]{\architectT}} &
			\ifthenelse{\lengthtest{\baselineT pt > 3599 pt}}{-}{\ifthenelse{\lengthtest{\baselineT pt < 0 pt}}{-}{\tablenum[round-mode=places, round-precision=2]{\darchT}}} &
			\tablenum[table-format=5,zero-decimal-to-integer]{\subsPi} &
			\ifthenelse{\lengthtest{\subsT pt > 3599 pt}}{\SI{>1}{\hour}}{\tablenum[round-mode=places, round-precision=2]{\subsT}} & 
			\tablenum[table-format=4,zero-decimal-to-integer]{\subslPi} &  
			{\bfseries \tablenum[round-mode=places, round-precision=2]{\subslT}} & 
			\ifthenelse{\lengthtest{\subsG pt < 0 pt}}{-}{\tablenum[round-mode=places, round-precision=2]{\dsubslT}}
			}
	\end{tabular}}\\\vspace{1mm}
	{\scriptsize $n$: Number of qubits \hspace*{0.3cm} \emph{$\vert G \vert$}: Gate count of $G$ \hspace*{0.3cm} $c$: Resulting cost (i.e., added gates to satisfy all coupling constraints) \\ \emph{$|\Pi|$}: Original number of considered permutations \hspace*{0.3cm} \emph{$|\Pi'|$}: Maximum number of reduced permutations\\ \emph{$t_\mathit{ref}$}: Runtime of JKQ QMAP~\cite{willeMappingQuantumCircuits2019} (with and without subgraphs) \hspace*{0.4cm} \emph{$t_\mathit{prop}$}: Runtime of proposed scheme (with and without subgraphs) \\ The optimal cost~$c$ has been achieved by \emph{all} approaches, independently of whether all permutations~$\Pi$ or just the limited set $\Pi'$ have been considered.}\vspace*{-0em}
\end{table*}

The observations and resulting strategies proposed above can, in general, be employed on top of any optimal method for the mapping problem (such as~\cite{hirataEfficientConversionQuantum2011,siraichiQubitAllocation2018,zhangTimeoptimalQubitMapping2021, tanOptimalLayoutSynthesis2020,dealmeidaFindingOptimalQubit2019,zhuExactQubitAllocation2020, willeMappingQuantumCircuits2019}). 
In order to experimentally evaluate their effect, we implemented the strategies proposed in~\autoref{sec:heuristics} on top of the quantum circuit mapping tool QMAP, which is based on the method proposed in~\cite{willeMappingQuantumCircuits2019} and publicly available at \url{https://github.com/iic-jku/qmap} as part of the open-source JKQ toolkit for quantum computing~\cite{willeJKQJKUTools2020}. 
This led to a version which computes the necessary permutations $\Pi'$ prior to the execution of the mapping, as well as a version that considers all possible \mbox{permutations $\Pi$}\footnote{For a comparison between optimal and heuristic approaches we refer to previous work, such as~\cite{willeMappingQuantumCircuits2019}, which shows that heuristics frequently stray far from the achievable optimum.}.

We further distinguished the evaluations between those that determine these permutations based on the complete coupling graph (and the corresponding strategy proposed in \autoref{sec:ideaA}) and those that determine these permutations only based on the necessary \textit{subgraphs} (and the corresponding strategy proposed in \autoref{sec:ideaB}).
All evaluations have been conducted on an AMD Ryzen 9 3900X processor with \SI{4.1}{\giga\hertz} and \SI{128}{\gibi\byte} of main memory running Ubuntu 20.04 using a hard timeout of \SI{1}{\hour}.
All results have been verified using the method provided in~\cite{burgholzerAdvancedEquivalenceChecking2021}.

\autoref{tbl:evaluation_results} provides the obtained results. 
Here, the first four columns identify the benchmark\footnote{As benchmarks we used instances that have been frequently used by related work in the past. All circuits have been mapped to the $5$-qubit, \mbox{T-shaped} IBMQ London architecture. In addition to that, the implementation of the approach is publicly available so that the interested reader can conduct further evaluations.}, the number of qubits $n$, the number of gates $|G|$ and the optimal mapping cost~$c$ (i.e., the additional number of gates needed to satisfy all coupling constraints). Note that the optimal mapping cost~$c$ has been achieved by \emph{all} approaches, independently of whether all permutations~$\Pi$ or just the limited set of permutations~$\Pi'$ have been considered.
Afterwards, we list for all approaches the total number of considered permutations ($|\Pi|$ in case of the original approach and $|\Pi'|$ in case of the proposed approach) as well as the respectively required runtimes in CPU seconds ($t_\mathit{ref}$ in case of the reference approach considering \emph{all} permutations~$\Pi$ and $t_\mathit{prop}$ in case of the proposed approach considering the limited number of permutations~$\Pi'$). 

First and foremost, the results confirmed that \emph{all} approaches yield circuits with the minimal mapping cost. This is perfectly in line with the theoretical discussion in \autoref{sec:theory} and confirms that, limiting the search space as proposed in this work, still guarantees optimal results.

At the same time, limiting the search space drastically reduces the complexity of the problem. The respective columns in \autoref{tbl:evaluation_results} denoted~$|\Pi|$ and~$|\Pi'|$ show the difference. For example, rather than \num{26040} permutations, only \num[group-minimum-digits=4]{3472} permutations need to be considered in case of benchmark \emph{4\_49\_16} when mapping to the IBMQ London architecture. 
In this particular case (and some others, as reported in \autoref{tbl:evaluation_results}), this makes the differences between  running into a timeout of \SI{1}{\hour} or being able to determine an optimal result in just some minutes. But also in the cases where the reference approach succeeds in obtaining a result within \SI{1}{\hour}, limiting the search space proves beneficial. In fact, for \emph{all} benchmarks substantial speed-ups can be observed. 
In the best case, speed-ups of up to three orders of magnitude are possible.
And, again, these improvements are possible while, at the same time, still obtaining optimal results. 

\section{Conclusions}
\label{sec:conclusions}

In this work, we proposed generic and \mbox{architecture-independent} strategies to limit the search space that needs to be considered when aiming for the determination of optimal results for the quantum circuit mapping problem. 
These strategies are motivated by observations showing that only a limited set of permutations in front of each gate needs to be considered---addressing the origin of the huge search space and complexity.
Theoretical considerations (based on group theory) back these observations and experimental evaluations confirm the resulting benefits: Limiting the search space as proposed in this work allows to drastically improve the performance of corresponding approaches (allowing to complete instances within minutes that ran into a timeout before or to achieve speed-ups of up to three orders of magnitude) while, at the same time, remaining optimal.

\section*{Acknowledgments}
This project has received funding from the European Research Council (ERC) under the European Union’s Horizon 2020 research and innovation programme (grant agreement No. 101001318).
It has partially been supported by the LIT Secure and Correct Systems Lab funded by the State of Upper Austria as well as by the BMK, BMDW, and the State of Upper Austria in the frame of the COMET program (managed by the FFG).

\printbibliography
\end{document}